\documentclass[aps,twocolumn,amsfonts,prl,nofootinbib,showpacs]{revtex4}
\usepackage{graphicx}

\def\L       {{\mathcal L}}
\def\D       {{\mathcal D}}

\def\pd      { \partial}
\def\V       {{\mathcal V}}

\def\L       {{\mathcal L}}

\def\pd { \partial}
\def\V       {{\mathcal V}}
\def\bra     {{\langle}}
\def\ket     {{\rangle}}
\def\a{\alpha}
\def\b{\beta}

\def\c{\chi}

\def\be{\begin{equation}}
\def\ee{\end{equation}}
\newcommand{\eea}{\end{eqnarray}}
\newcommand{\bea}{\begin{eqnarray}}

\def\d{\delta}
\def\l{\lambda}
\def\d{\delta}

\def\l{\lambda}

\def \p { \phi}

\def\pdbk{\bra \phi^\dag \ket}
\def\pbk{\bra \phi \ket}

\def\V2{\mu^2 \pdbk \pbk + \lambda (\pdbk \pbk )^2 }
\usepackage{bm}
\usepackage{epsfig}

\begin{document}

\title{A Phase Transition in U(1) Configuration Space: Oscillons as Remnants of Vortex-Antivortex Annihilation}


\author{M.~Gleiser}
\email{gleiser@dartmouth.edu}
\author{J.~Thorarinson}
\email{thorvaldur@dartmouth.edu}

\affiliation{Department of Physics and Astronomy, Dartmouth College,
Hanover, NH  03755, USA\\ \\}

\begin{abstract}
We show that the low-momentum scattering of vortex-antivortex pairs can lead to very long-lived oscillon states in 2d Abelian Higgs models. The emergence of oscillons is controlled by the ratio of scalar and vector field masses, $\beta=(m_s/m_v)^2$ and can be described as a phase transition in field configuration space with critical value $\beta_c\simeq 0.13(6)\pm 2 $: only models with $\beta<\beta_c$ lead to oscillon-like remnants.  The critical behavior of the system obeys a power law $O(\beta)\sim |\beta-\beta_c|^o$, where $O$ is an order parameter indicating the presence of oscillons and $o = 0.2(2)\pm 2 $ is the critical exponent.

\end{abstract}

\maketitle

\noindent
{\it Introduction.} In relativistic and nonrelativistic field theories, models that exhibit spontaneous symmetry breaking support a plethora of topologically-stable field configurations or defects \cite{Shellard-Vilenkin}, static solutions of the nonlinear equations of motion with properties determined by the topology of the vacuum manifold. One of the simplest examples is the kink in 1d $\lambda\phi^4$ models, the result of a discrete symmetry breaking: the vacuum has two disconnected points and the kink interpolates between them. Once formed, kinks cannot be destroyed. Unless, of course, they meet an antikink. Low-momentum kink-antikink scattering leads to the formation of breathers, long-lived time-dependent bound states characterized by large-amplitude oscillations of the scalar field \cite{Rajamaran}.  At larger velocities, breathers have been shown to form only at certain windows or resonances \cite{Campbell}.  Although $\lambda\phi^4$ breathers have not been seen to decay in numerical studies \cite{Campbell}, they may decay via nonperturbative effects \cite{Segur-Kruskal}, radiating their energy to spatial infinity.

In 2d, the simplest stable topological defect with localized energy is the Nielsen-Olesen vortex found in Abelian Higgs models (AHMs) \cite{Shellard-Vilenkin}. These vortices are of great interest in many areas of physics, e.g., as prototypes for studying nonperturbative effects in the Standard Model and its extensions, or, in the nonrelativistic limit, as Abrikosov vortices in the Landau-Ginzburg theory of superconductivity \cite{superconductivity}. Vortex-like solutions may also have a variety of cosmological roles, being formed during a phase transition at the GUT scale or thereafter \cite{Shellard-Vilenkin,cosmology} or during reheating after inflation \cite{reheating}.

Inspired by the existence of long-lived breathers in 1d $\lambda\phi^4$ models, in this Letter we investigate the low-momentum scattering of vortices and antivortices in AHMs. Vortices and antivortices 
carry equal and opposite topological charge. This allows for their annihilation. The only adjustable parameter in the model is $\beta\equiv \lambda/2 g^2$, the ratio of squared masses of the scalar and vector fields. We will show that for $\beta$ smaller than a critical value $\beta_c$, vortex-antivortex (henceforth vav) annihilation does give rise to higher-dimensional breather-like configurations known as oscillons \cite{oscillons}, also characterized by long-lived oscillations of the Higgs field magnitude. They were found in 2d \cite{2dosc}, 3d \cite {3dosc} and higher-dimensional real scalar field models \cite{dosc}. Their properties have been studied in great detail \cite{recentosc}. Long-lived oscillatory solutions have also been found in the decay of sphalerons in the 1d AHM \cite{1dsphalerons}. Recently, oscillons were found in 3d gauged SU(2) models \cite{Farhi} and gauged SU(2)xU(1) models \cite{Graham} when the Higgs mass is twice the $W^{\pm}$ boson mass. 

In the above references, the procedure was to solve the relevant equations of motion assuming or not spherical symmetry and using an initial profile that approximates the oscillon solution. So long as certain conditions are satisfied \cite{dosc}, the system relaxes into oscillons as it evolves in time. Here, we show that oscillons may also emerge dynamically, as remnants of vav scattering. (Previous attempts to find oscillons in vortex-vortex scattering were unsuccessful \cite{Moriarty:1988fx}.) Our finding complements recent results where oscillons were shown to form after a rapid quench in real scalar field models with symmetric \cite{Gleiser-Howell1} and asymmetric \cite{Gleiser-Howell2} potentials. The fact that oscillons may emerge dynamically raises their stakes considerably: they can play a crucial role during spontaneous symmetry-breaking in a variety of physical systems and situations, from the early universe to high energy collisions, and possibly in superfluids and superconductors.\\


\noindent
{\it The Model.} The Abelian Higgs Lagrangian density is:
\be
\label{Lagden}
 \L=\D_\mu \phi^\dag \D^\mu \phi-\frac{1}{4}F^{\mu \nu}F_{\mu
\nu}- \frac{\l}{4}(\phi^\dag\phi-\eta^2)^2, 
\ee 
where $\D_\mu=\pd_\mu-igA_\mu$. The scalar and vector masses are, respectively, $m_s=\sqrt{\lambda}\eta$ and $m_v=\sqrt{2}g\eta$. Their ratio defines the parameter $\beta=(m_s/m_v)^2=\lambda/2g^2$. In the nonrelativistic limit, $\beta=1$ defines the boundary between Type I ($\beta <1$) and Type II ($\beta>1$) superconductors.
With the scaling $A_\mu\rightarrow
\eta^{-1}A_\mu$, $\phi\rightarrow \eta^{-1}\phi$ and $x\rightarrow
\eta g x$, the  energy is $E= \eta^2 \int_{M} d^2x \mathcal H$, where the Hamiltonian
density in the temporal gauge $A_0=0$ is:
 \be 
 \mathcal H = \frac{1}{2}(E^2 +
B^2) +\pd_t \phi^\dag \pd_t \phi + \D_i \phi^\dag \cdot \D_i
\phi + \frac{\beta}{2}(\phi^\dag\phi-1)^2. 
\ee

From eq. \ref{Lagden}, one derives the equations of motion
\be
\label{eom}
\D_\mu\D^\mu \phi = -\frac{\l}{2}\phi\left (|\phi|^2-\eta^2 \right );
\pd_\mu F^{\mu \nu}=j^\nu=2 g{\rm Im}\left\{\phi^\dag \D^\nu \phi
\right\}.
\ee
Nielsen-Olesen vortices are solutions satisfying $\phi({\bf r})= e^{in\theta}f(r)$ and $A({\bf r})_{\theta}=-(n/gr)\alpha(r)$, with boundary conditions at spatial infinity [$r=(x^2+y^2)^{1/2}\rightarrow \infty$]
$f(r),\alpha(r)\rightarrow 1$. $n$ is the winding number. At the vortex center, $f(0)=\alpha(0)=0$. Vortices have a quantized magnetic flux $\Phi_B=2\pi n/g$. For a vav pair, the net flux is zero. Quantities are measured in units of $\eta=1$.\\


\noindent
{\it Lattice Implementation.} We work on the temporal gauge $A_0=0$ and follow the Hamiltonian implementation for lattice gauge theories described in refs. \cite{Moriarty:1988fx,SmitBook}.
The discrete Hamiltonian is:  
\be 
H=\sum_x \mathcal H(\pi^\a_x,\phi^\a_x)  \d x^2, 
\ee 
where the density is: 
\bea 
& \mathcal H &= \pi_x^\dagger \pi_x +\vec \D
\p_x^\dagger \cdot \vec\D \p_x
+\frac{1}{2}\vec{E}_{x} \cdot \vec{E}_{x} \\ \nonumber
&+& \frac{1}{2 \d x^2} \sum_{i\neq j}(A^{ i}_{ x+ j}- A^{ i}_{ x}
-A^{ j}_{ x + i }+A^{ j}_{ x} )^2 +V(\phi_x^\dag\phi_x),
\eea 
and the lattice covariant derivative is $\D_\mu \p=(e^{-i g\d x A^\mu_x }
\p_{x+\mu}-\p_x) \d x^{-1}$. The Hamiltonian density is invariant under that lattice gauge transformations:
\bea 
\p_{{x}} &\rightarrow&e^{i \c_{{x}}} \p_{{x}};~ \pi_{{x}} \rightarrow  e^{i \c_{{x}}} \pi_{{x}}\\ \nonumber
A^\mu_{{x}} &\rightarrow&A^\mu_{{x}} +
(\c_{{x}+{\mu}}
- \c_{{x}})/(g\d x); ~ E^\mu_{{x}} \rightarrow E^\mu_{{x}}. \nonumber
\eea
The equations of motion are then:
\bea
\pd_t\pi^\a_x&=&-\pd_{\p_x} \sum_x \mathcal H(\pi^\a_x,\phi^\a_x); \\ \nonumber
\pd_t\p^\a_x&=&\pd_{\pi_x} \sum_x \mathcal H(\pi^\a_x,\phi^\a_x)
\eea
for each of the fields $\pi^\a=\{\pi,E_i\}$ and $\p^\a=\{\phi,A_i\}$. The temporal evolution follows a second-order leapfrog scheme with $\d t<\d x /\sqrt{2} $.  Provided that the initial conditions satisfy the lattice Gauss's law constraint,
\be \frac{1}{\d x} \sum_i (E^i_{x-i}-E^i_x)=i(\pi_x\p_x^\dagger -\pi_x^\dagger \p_x),
\ee  
it will remain satisfied during the evolution. This can be implemented by, e.g., setting all temporal derivatives to zero as the simulation starts. Energy is conserved to $O(\d t^2)$. We used $\d t=0.02$ and $\d x = 0.2$ and checked that the results are consistent for $\d x = \{0.1\leftrightarrow 0.3 \}$. Larger values compromise the accuracy of the results.

Since we use periodic boundary conditions, to prevent energy from the initial condition to come back to the configuration we use a lattice much larger then the region occupied by the vav pair.  However, as oscillons are very long-lived, we also use an adiabatic absorbing wall \cite{2dosc} implemented by adding gauge invariant $\gamma(x,y) \pd_t \phi $ terms to the evolution equation, where $\gamma(x,y)$ has support only on a band near the lattice boundary and slowly increases from $0\rightarrow 0.1$.\\


\noindent
{\it Vortex Antivortex Scattering: Results.} We set the initial conditions by using an {\it ansatz} that approximates a vav pair separated by an initial distance $\delta_0$. This is then allowed to evolve in time under the equations of motion, eq. \ref{eom}, at high viscosity until it settles into the vav configuration. At large $\delta_0$ and for $\beta \leq 1$ (the case of interest to us) we computed numerically the total energy of a vortex, finding a good fit for
\be
E_v =2\pi \beta^{1/5},
\ee
which agrees near the region $\beta\sim 1$ with the result of \cite{Jacobs-Rebbi}. Also, the  contribution in gauge fields to the vortex energy, $E_g=(E^2+B^2)/2$, can be fitted as
\be
E_g = \frac{\pi}{\sqrt{6}}\beta^{1/4}.
\ee
After the initial {\it ansatz} settles into a vav pair, we let them approach until their centers reach the distance $\delta_i=4m_s^{-1}$, large enough to avoid any tail overlap \cite{Jacobs-Rebbi}. At this point, we set the viscosity to zero and let the pair attract and scatter. We then repeat the experiment for different values of $\beta$.

In Fig. 1 we show a sequence of snapshots of an annihilation process for $\beta=0.08$, where an oscillon is formed. We plot both the amplitude $|\phi|^2$ and the magnetic field $B_z$. One can see the outgoing wavefront of radiation (center) as the vortices coalesce into the oscillon and its oscillating dipole magnetic field (right).

\begin{figure}
\includegraphics[scale=.5,angle=90]{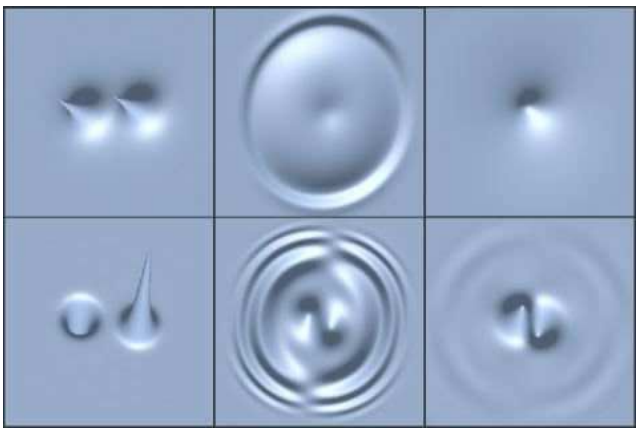}
 \caption{\label{snapshots} Snapshots of $|\phi|^2$ (top)  and  $B_z$ (bottom) for a configuration with $\beta=0.08$. Time proceeds from the initial vav pair on the left to the oscillon on the right.  } 
\end{figure}

In Fig. 2 we plot the associated energy ($\cal{H}$) and Chern-Simons ($n_{CS}$) densities, where $n_{CS}=(4\pi)^{-1}\varepsilon_{\alpha\beta\gamma}A^{\gamma}F^{\alpha\beta}$. Note that oscillons have a distinctive signature and thus could have an interesting role in theories with current anomalies of the form $\pd_\mu J^\mu_B\propto F\tilde{F}$ leading to baryogenesis, including during postinflationary reheating \cite{preheating}, a possibility we are exploring.

\begin{figure}
\includegraphics[scale=.5,angle=90]{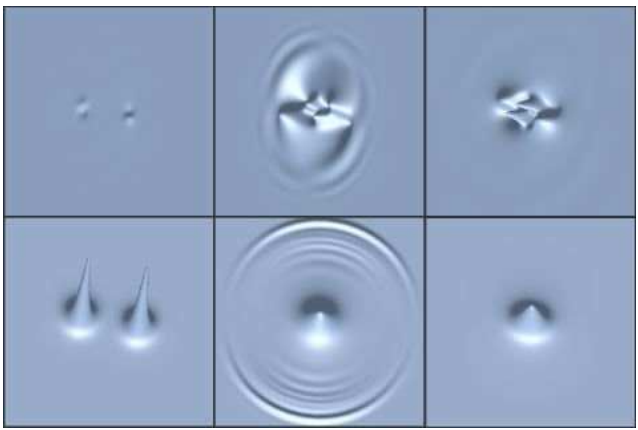}
 \caption{\label{Eden} Chern-Simons density (top) and energy density (bottom) for the same snapshots of Fig. 1 } 
\end{figure}

Vortex-antivortex scattering can lead to two possible outcomes, depending on the value of $\beta$: either the vav pair quickly annihilates, $(+) + (-)\rightarrow 0$, or it settles into an intermediate, long-lived oscillatory state before radiating its energy to spatial infinity,  $(+) + (-)\rightarrow (+-) \rightarrow 0$. In order to investigate the process quantitatively, we measured the energy in a disk ($E_{\rm disk}$) surrounding the scattering process. The results are shown in Figs. 3 and 4. In Fig. 3 we show the energies in a disk of radius $r=15$ for different values of $\beta$. In Fig. 4 we show the total energy and the associated energy in gauge fields for $\beta=0.052$. The inset shows that there are two characteristic frequencies, set by the scalar and gauge field masses, $m_s$ and $m_v$. For $\beta=0.052$, one obtains $m_v\sim 4.4 m_s$.

\begin{figure}
\includegraphics[scale=.6,angle=0]{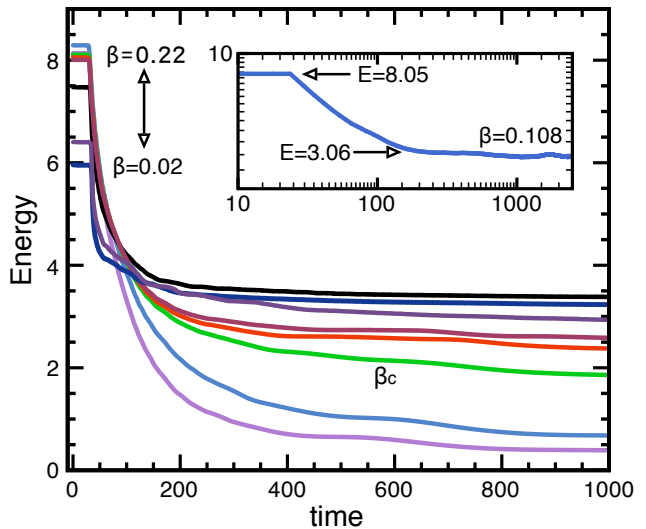}
 \caption{\label{Eplot} Energy within a disk of radius $r=15$ surrounding the initial vav pair for various values of $\beta$ on a $100^2$ lattice. The range of $\beta$ is indicated on the top left. There are two regimes: for $\beta\geq \beta_c\simeq 0.136$, the pair quickly annihilates, radiating its energy to spatial infinity. For $\beta<\beta_c$, the pair sheds some of its initial energy and settles into a long-lived intermediate state, where almost no energy is radiated. This state is the U(1) Abelian Higgs oscillon. Inset: change in energy (E) radiated as a function of $\ln t$ for $\beta=0.108$. Note the ``hockey-stick'' shape. The arrow denotes the oscillon's energy, $E_{\rm osc}$.} 
\end{figure}

\begin{figure}
\includegraphics[scale=.6,angle=0]{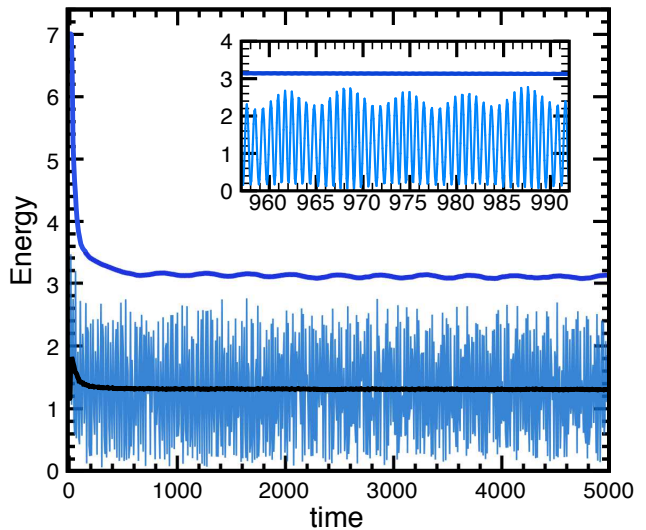}
 \caption{\label{fig:q6vor2osc} Top curve: total energy for $\beta=0.052$ in a disk of radius $r=15$ surrounding the initial vav pair. Lower curve:  energy in the gauge fields. Black line is the running average of the gauge energy. The inset shows the total energy and the energy in gauge fields for a small time interval. The two characteristic frequencies satisfy $m_s\sim\sqrt{\b}m_v$. }
\end{figure}

By inspecting Fig. \ref{Eplot}, we identify an oscillon whenever the energy develops a plateau. To read off a value of the plateau energy we examine the energy vs. $\ln t$ plot. When oscillons are present, the curve assumes a ``hockey stick'' shape. We read off the oscillon's energy ($E_{\rm osc}$) as the value of $E_{\rm disk}$ at the location of the elbow. [Arrow  in inset in Fig. \ref{Eplot}.] 
We have observed that oscillon lifetimes are all in excess of $\tau_{\rm osc}\geq 10^{(4-5)}$. For example, for $\beta=0.098$,  $\tau_{\rm osc}\geq 1.1\times 10^5$. We leave a detailed study of their extreme longevity for future work. 

We identify $\beta_c=0.13(6)\pm 2$ as the critical value separating the two regimes, as it exhibits the critical slowing down typical of continuous phase transitions. For $\beta>0.2$, well outside the critical region, no oscillons form: vav pairs radiate their initial energy to spatial infinity so that by $t\sim 500$,  $E_{\rm disk}\lesssim 0.4\ll E_{\rm osc}$. 
[We neglect here some small remnant energy $E_{\rm disk} < 0.1-0.2$, that persists long beyond the oscillon's demise.]

\begin{figure}
\includegraphics[scale=.575,angle=0]{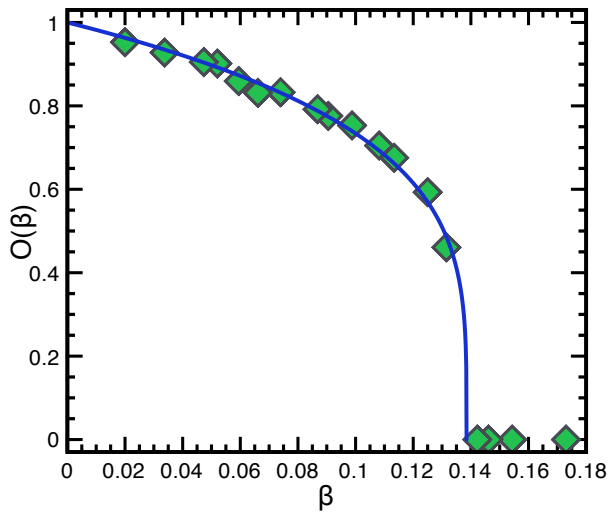}
 \caption{\label{phasediag} Phase diagram for the vav $\rightarrow$ oscillon transition. Plotted is the order parameter $O(\b)\equiv E_{\rm osc}/E_v$ as a function of $\beta$. The diamonds are numerical results, while the continuous line is the fit of eq. \ref{phasetran}.}
\end{figure}

The quantity $E_{\rm osc}/E_v$ naturally lends itself as an effective order parameter describing the transition from vav pairs to oscillons in field configuration space. In Fig. \ref{phasediag} we plot $E_{\rm osc}/E_v$ vs. $\beta$. The curve is well fitted for 
\be
\label{phasetran}
E_{\rm osc}/E_v \sim (|\beta-\beta_c |)^{0.2(2)\pm 2}.
\ee
Near criticality the exponent $0.2$ will have small corrections \cite{phase transitions}. Given the long integration times needed to extract a more precise value for the energy of the oscillons around $\beta_c$ (akin to critical slowing down, as can be seen in Fig. \ref{Eplot}), we did not attempt to find these corrections here. From the theory of continuous phase transitions, the critical point is related to infrared divergences. In fact, the effective size of the oscillon increases near $\beta_c$: disks of larger radii are needed to capture the full energy of the configuration. The associated decrease in the effective scalar field mass near the core [$\sim (A^2-\beta)^{1/2}$] is the result of screening by the gauge field. Above criticality, the mass at the core becomes positive, eventually matching the asymptotic vacuum as the vav annihilates.

Our results show that the annihilation of a vav pair in the 2d AHM can be described as a phase transition in configuration space, with complete annihilation corresponding to the ``symmetric'' phase $E_{\rm osc}/E_v \rightarrow 0$, and oscillons corresponding to long-lived ``broken-symmetric'' states, with the associated order parameter attaining a nonzero value. 

It will be interesting to investigate if oscillons can exist in superconductors, how their lifetimes vary with $\beta$, and if they can be created in non-Abelian models via, for example, low momentum monopole-antimonopole scattering.

We thank Noah Graham for stimulating discussions.

\end{document}